\documentclass{article}

\usepackage[final,nonatbib]{nature_preprint}

\usepackage[utf8]{inputenc} 
\usepackage[T1]{fontenc}    
\usepackage{hyperref}       
\usepackage{url}            
\usepackage{booktabs}       
\usepackage{amsfonts}       
\usepackage{nicefrac}       
\usepackage{microtype}      

\usepackage{subcaption, booktabs}
\usepackage[labelfont=bf]{caption}

\usepackage[pdftex]{graphicx}
\usepackage[numbers]{natbib}

\usepackage{fixltx2e}
\usepackage{algorithm}

\usepackage{gensymb}
\usepackage{import}
\usepackage{cleveref}
\graphicspath{ {./figures/} }
\usepackage{setspace}
\usepackage{authblk}

\title{\LARGE Robust breast cancer detection in mammography and digital breast tomosynthesis using annotation-efficient deep learning approach\vspace{-10pt}}

\author[1]{William Lotter}
\author[1]{Abdul Rahman Diab*}
\author[1]{Bryan Haslam*}
\author[1]{Jiye G. Kim*}
\author[1]{Giorgia Grisot}
\author[1]{Eric Wu}
\author[1]{Kevin Wu}
\author[1]{Jorge Onieva Onieva}
\author[2]{Jerrold L. Boxerman}
\author[3]{Meiyun Wang}
\author[4]{Mack Bandler}
\author[5]{Gopal Vijayaraghavan}
\author[1]{A. Gregory Sorensen\vspace{-3pt}}

\affil[1]{DeepHealth, Inc.}
\affil[2]{Rhode Island Hospital \& Brown University}
\affil[3]{Henan Provincial People's Hospital}
\affil[4]{Medford Radiology Group}
\affil[5]{University of Massachusetts Medical School}
\begin{document}
\maketitle
\let\thefootnote\relax\footnote{* Denotes equal contribution.
Address correspondence to: lotter@deep.health}

\vspace{-35pt}

\textbf{Breast cancer remains a global challenge, causing over 1 million deaths globally in 2018~\cite{Bray2018}. To achieve earlier breast cancer detection, screening x-ray mammography is recommended by health organizations worldwide and has been estimated to decrease breast cancer mortality by 20-40\%~\cite{Berry2005, Seely2018}. Nevertheless, significant false positive and false negative rates, as well as high interpretation costs, leave opportunities for improving quality and access~\cite{Majid2003, Rosenberg2006}. To address these limitations, there has been much recent interest in applying deep learning to mammography~\cite{Yala2019a, Yala2019, Conant2019, Rodriguez-Ruiz2019a, Wu2019, ribli_detecting_2018, Kooi2017, geras2017highresolution, Lotter2017}; however, obtaining large amounts of annotated data poses a challenge for training deep learning models for this purpose, as does ensuring generalization beyond the populations represented in the training dataset. Here, we present an annotation-efficient deep learning approach that 1) achieves state-of-the-art performance in mammogram classification, 2) successfully extends to digital breast tomosynthesis (DBT; ``3D mammography''), 3) detects cancers in clinically-negative prior mammograms of cancer patients, 4) generalizes well to a population with low screening rates, and 5) outperforms five-out-of-five full-time breast imaging specialists by improving absolute sensitivity by an average of 14\%. Our results demonstrate promise towards software that can improve the accuracy of and access to screening mammography worldwide.}
\\

\vspace{-5pt}
\indent While specific guidelines vary~\cite{Siu2016, Oeffinger2015, Mainiero2017, NHSBSP2016, WHO2014}, x-ray mammography has been the gold standard for breast cancer screening for decades. 
Mammogram acquisition technology has seen a number of improvements over this period, including the advent of DBT~\cite{Kopans_2014}, yet studies estimate that indications of a detected cancer’s presence are visible 20-60\% of the time in earlier exams that were interpreted as normal~\cite{Saarenmaa1999, Ikeda2003, Hoff2011}. 
Simultaneously, significant false positive rates lead to patient stress and expensive follow-up procedures. Mammogram interpretation is particularly difficult because abnormalities are often indicated by small, subtle features, and malignancies are only present in approximately 0.5\% of screened women. These challenges are exacerbated by the high volume of mammograms (39 million per year in the United States alone~\cite{mqsa_national_statistics}) and the additional time required to interpret DBT. 

\indent Given these challenges, there have been many efforts in developing computer-aided diagnosis (CAD) software to assist radiologists in interpreting mammograms. The motivation behind initial versions of CAD arose from the reasoning that - even if the standalone performance of a CAD system was inferior to expert humans - its use could still result in improved sensitivity when used as a ``second look'' tool. 
While early evidence supported this claim~\cite{Freer_2001}, the practical effectiveness of traditional CAD has been questioned, and several studies have shown that it can lead to drawbacks such as higher false positive rates and increased interpretation time~\cite{Fenton2007, Lehman2016, Henriksen2019, Tchou_2010}.

\indent A potential reason for the limited accuracy of traditional CAD software is that it relied on hand-engineered features. Deep learning, which has been shown to outperform feature engineering in many computer vision problems, relies instead on learning the features and classification decisions end-to-end. Recent applications of deep learning to mammography have shown great promise~\cite{Yala2019a, Yala2019, Conant2019, Rodriguez-Ruiz2019a, Wu2019}, and several commercial products have received FDA clearance as aids for human interpretation~\cite{Conant2019, Rodriguez-Ruiz2019b, curemetrix_FDA}. 
Despite this progress, there is still significant room for improvement, particularly in developing a deep learning approach that a) performs beyond average radiologist levels while b) generalizing to populations and data sources not used for model training and that c) extends to both 2D and 3D mammography.
As performance metrics can vary greatly depending on the tested dataset and ground-truth criteria, comparing directly to human experts is essential for the most meaningful benchmarking. Testing generalization to populations and datasets not used for model training is also imperative, as deep learning can be highly sensitive to input statistics~\cite{kim_design_2019, pmlr-v81-buolamwini18a, Szegedy2013, zech2018variable}. 
Lastly, developing an effective DBT model is increasingly important given the technology's growing use~\cite{mqsa_national_statistics} and the sparsity of deep learning solutions that leverage the advantages of 3D mammography.

\indent Both data-related and algorithmic challenges contribute to the difficulty of developing deep learning solutions that achieve the aforementioned goals. 
Deep learning models generally perform best when trained on large amounts of data, especially highly-annotated data, and obtaining either for mammography can be a formidable task.
The two most prominent publicly accessible datasets for 2D mammography are the US-based Digital Database of Screening Mammography (DDSM)~\cite{Heath2000}, which consists of digitized film exams with radiologist-drawn segmentation maps for detected lesions, and the UK-based Optimam Mammography Image Database (OMI-DB)~\cite{omidb}, which contains digital mammograms with a mix of strongly-labeled (in the form of bounding boxes) and weakly-labeled (breast-level) data. We are not aware of any public datasets for DBT. Strongly-labeled data, where lesion localization information is available, is particularly valuable for mammography given its ``needle-in-a-haystack'' nature. This is especially true for DBT, which can contain over 100 times as many images (or ``slices'') as DM and on which malignant features are often only visible in a few of the slices. From a machine learning perspective, this combination of large data size and relatively small findings can lead to significant overfitting. Strongly-labeled data can help mitigate this overfitting, but such data can be costly or impractical to collect, as well as noisy due to annotator variability. Additional challenges include the severe class imbalance of screening mammography (99.5\% of cases are non-cancerous) and the high resolution of mammographic images, which can be 10-20 times the resolution typically used for approaches on natural images.

\begin{figure}[h]
	\begin{center}
		\includegraphics[width=1.0\textwidth]{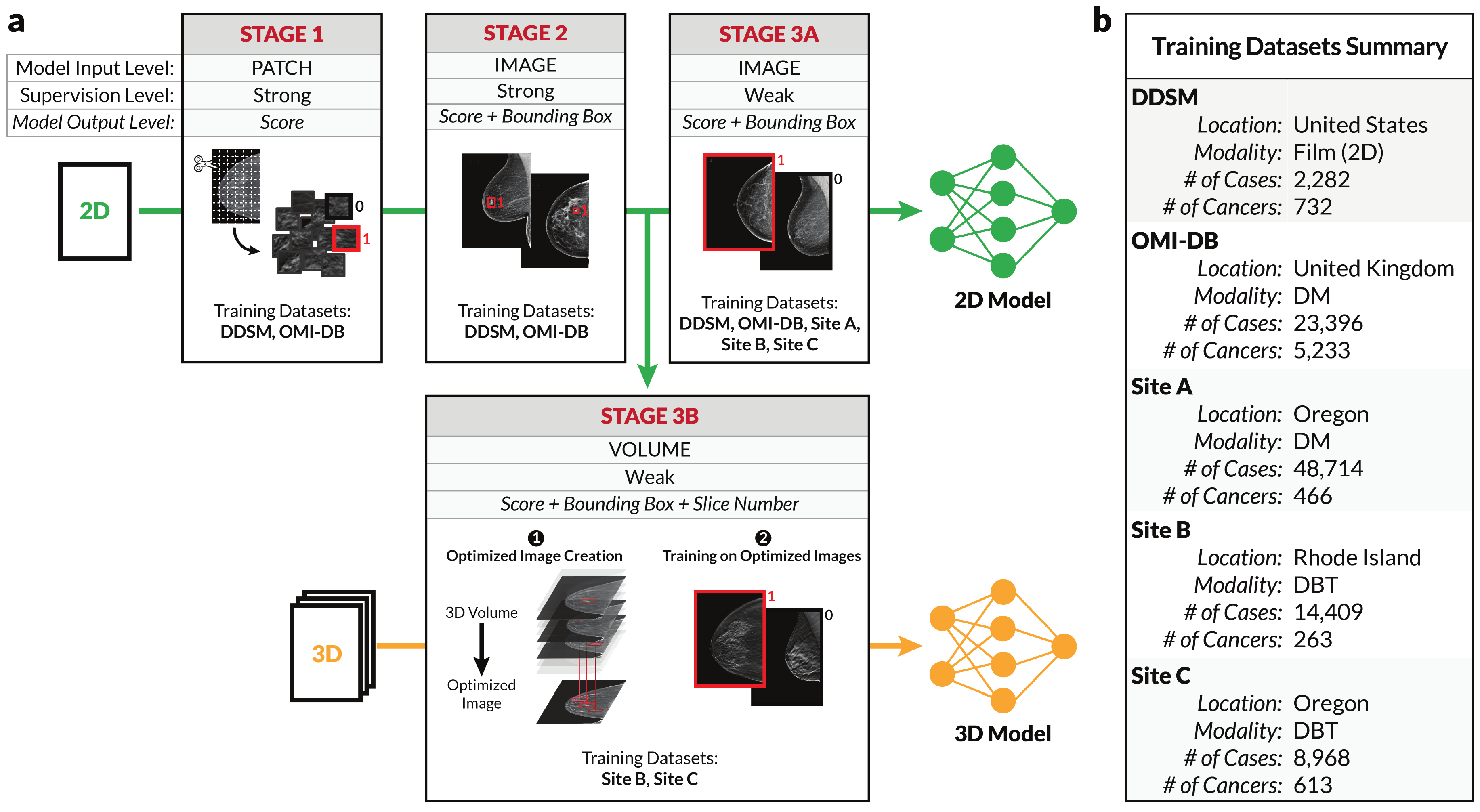}
	\end{center}
	\vspace{-10pt}
	\caption{\textbf{Model training approach.} \textbf{a)} To effectively leverage both strongly and weakly-labeled data while mitigating overfitting, we train our deep learning models in a series of stages. Stage 1 consists of patch-level classification using cropped image patches from 2D mammograms~\cite{Lotter2017}. In Stage 2, the model trained in Stage 1 is used to initialize the feature backbone of a detection-based model. The detection model, which outputs bounding boxes with corresponding classification scores, is then trained end-to-end in a strongly-supervised manner on full images. Stage 3 consists of weakly-supervised training, for both 2D and 3D mammography. For 2D (Stage 3a), the detection network is trained for binary classification in an end-to-end, multiple-instance learning (MIL) fashion where an image-level score is computed as a maximum over bounding box scores. For 3D (Stage 3b), the model from Stage 2 is used to condense each DBT stack into an optimized 2D image by evaluating the DBT slices and extracting the most suspicious regions of interest at each x-y spatial location. The model is then trained on the optimized images using the same MIL approach as Stage 3a. \textbf{b)} Summary of training datasets.} 
	\label{methods}
\end{figure}

\indent Here, we have taken steps to address both the data and algorithmic challenges of deep learning applied to mammography.
To address data scarcity, we have assembled three additional training datasets, focusing especially on DBT.
To leverage this data in addition to public data, we have developed an algorithmic approach that effectively makes use of both strongly and weakly-labeled data by training a core model on a series of increasingly difficult tasks.



\indent Figure~\ref{methods} details our model training pipeline. Stage 1 consists of patch-level classification~\cite{Lotter2017}. Using the strongly-labeled data from OMI-DB and DDSM, we generate a large number of cropped image patches, where each patch is given a label according to whether or not it contains a lesion. We then train a popular convolutional neural network, ResNet-50~\cite{He2016}, to classify these patches. We next use the patch-trained ResNet-50 to initialize the backbone of a fully-convolutional, detection-based model, RetinaNet~\cite{retinanet}, which is then trained on full images in Stage 2. In the RetinaNet model, the ResNet-50 is used to create a feature pyramid that contains features extracted at different spatial scales over the entire image. The feature pyramid becomes the input to a detection module, which ultimately outputs bounding boxes at different locations across the image. 
Each bounding box is assigned a classification score, trained here to indicate the likelihood that the enclosed region represents a malignancy. 
In this second training stage, we again train in a strongly-supervised fashion using the strongly-labeled data from OMI-DB and DDSM. 

\indent To take advantage of all of the training data available, the final phase of our training pipeline consists of weakly-supervised training. 
We conduct this weakly-supervised training separately for 2D (Stage 3A) and 3D (Stage 3B) data. In both cases, we build upon the strongly-supervised training by using the same core RetinaNet architecture, initializing with the weights from Stage 2. For 2D inputs, we use a multiple-instance learning (MIL) formulation where a maximum is computed over all of the bounding box classification scores.
This results in a single, interpretable score for the entire image, corresponding to the most suspicious region of interest (ROI).
The resulting model remains fully-differentiable, allowing end-to-end training using binary cancer/no-cancer labels.
Importantly, even though the model at this stage is only trained with image-level labels, it retains its localization-based explainability, mitigating the ``black-box'' nature of standard classification models.

\indent For DBT, our approach to weakly-supervised training is motivated by the value of DBT in providing an optimally-angled view into a lesion that could otherwise be obscured by overlapping tissue, and by the similarity between DBT and DM images, which suggests the applicability of transfer learning. Furthermore, we especially consider that the aggregate nature of 2D mammography can help reduce overfitting compared to training end-to-end on a large DBT volume. With these intuitions in mind, we train on the DBT volumes in two steps. In the first step, the most suspicious regions in the DBT volume are condensed into an optimized image. To achieve this, the 2D strongly-supervised model separately evaluates the slices in the DBT stack, producing a set of bounding boxes at each spatial location for each evaluated slice. The bounding boxes over all of the slices are then aggregated and filtered using non-maximum suppression (NMS), which filters overlapping bounding boxes by their classification scores, retaining the highest scoring non-overlapping boxes. This results in a final set of bounding boxes wherein each box has the highest score at its spatial location over all of the evaluated slices. The image patches defined by the filtered bounding boxes are then collapsed into a single 2D array representing an image optimized for further model training. 
The strongly-supervised 2D RetinaNet model is then trained on these optimized images using the same MIL formulation described above.

\indent Figure~\ref{methods}b summarizes the data used to train our models. In addition to OMI-DB and DDSM, we use datasets collected from three clinical sites in the United States. During training, we implement class balancing by randomly sampling a cancer or non-cancer example with equal probability at each training iteration. 
This class balancing is implemented within datasets as well to prevent the model from learning biases in the different proportions of cancers across datasets.
Final model selection is based on performance on a held-out validation set, with cross-validation performed at the patient level. 

\indent To assess the performance of our deep learning approach, we conducted a reader study using screening DM cases retrospectively collected from a regional health system located in a different US state than the other sources of training data. No data from this site was ever used for model training or selection. Five radiologists who are full-time fellowship-trained breast imagers participated in the study.

\indent The first component of the reader study consisted of comparing the performance of the radiologists to the standalone performance of the deep learning model on ``index'' cancer mammograms. 
Index exams are the standard cases included in most reader studies as ``cancers'' and correspond to the screening exam from the year in which a given cancer was detected. 

\begin{figure}[t]
	\begin{center}
		\includegraphics[width=1.0\textwidth]{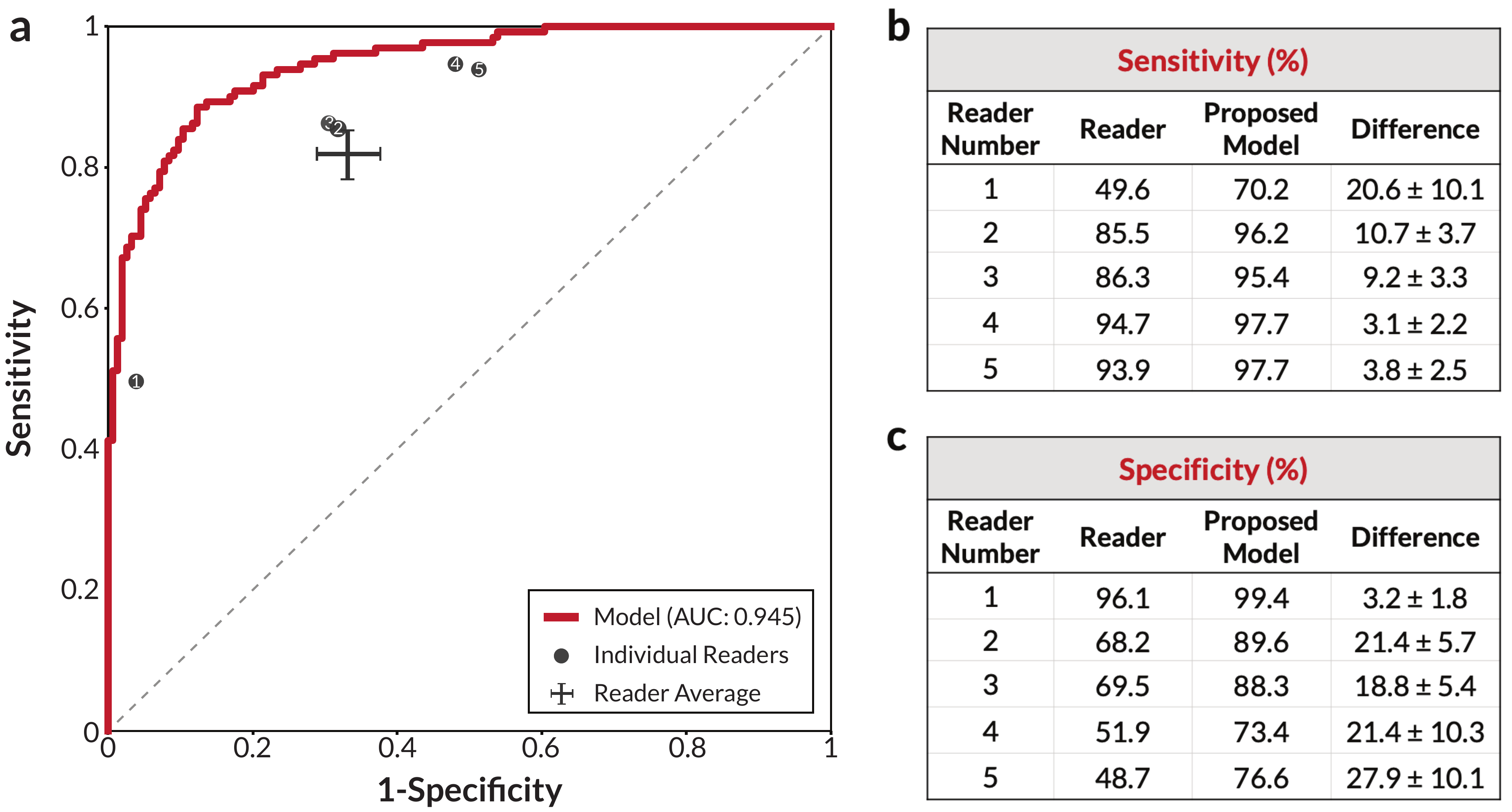}
	\end{center}
	\vspace{-3pt}
	\caption{\textbf{Reader study results -- index cancer exams}. \textbf{a)} The proposed deep learning model outperformed all five radiologists on the set of 131 index cancer exams and 154 confirmed negatives. Each data point represents a single reader, and the ROC curve represents the performance of the deep learning model. The cross corresponds to the mean radiologist performance with the lengths of the cross indicating 95\% confidence intervals for mean sensitivity and specificity. 
	\textbf{b)} Sensitivity of each reader and the corresponding sensitivity of the proposed model at a specificity chosen to match each reader. 
	\textbf{c)} Specificity of each reader and the corresponding specificity of the proposed model at a sensitivity chosen to match each reader.
	For \textbf{b)} and \textbf{c)}, the standard deviation of the model minus reader difference was calculated using 10K bootstrap samples.\vspace{-10pt}} 
	\label{gopal_index}
\end{figure}

\indent Figure~\ref{gopal_index} summarizes the results of the first component of the reader study. A receiver operating characteristic (ROC) plot is displayed, comparing the performances of the readers and our deep learning model on the set of 131 index cancer exams and 154 confirmed negatives (a negative screening exam followed by another negative exam at the next screen). Each point represents the sensitivity and specificity of one human reader, and the continuous ROC curve is calculated from the deep learning model’s output scores. Each reader's point falls below the deep learning model’s ROC curve, indicating that the model outperformed all five full-time breast imagers. At the average reader specificity, the model achieved an absolute increase in sensitivity of 14.2\% (95\% confidence interval (CI): 9.2-18.5\%; $p < 0.0001$). At the average reader sensitivity, the model achieved an absolute increase in specificity of 24.0\% (95\% CI: 17.4-30.4\%; $p < 0.0001$). Additionally, the model outperformed every combination of the readers (Extended Data Fig.~\ref{panel_figure}). The model’s area-under-the-curve (AUC) also exceeds that of recently-published models~\cite{Wu2019, Wu2019ValidationRates, Yala2019} (Extended Data Table~\ref{model_comparison_table}).

\indent The second component of our reader study involved the clinically-negative prior exams of the same cancer patients, which we denote as ``pre-index'' exams. These mammograms were obtained 12 to 24 months prior to the index exams of the same patient and were not flagged for further workup at the time of initial interpretation. The pre-index exams can largely be thought of as challenging false negatives, since it is estimated that breast cancer typically exists more than 3 years prior to detection by mammography~\cite{Hart1998, Weedon-Fekjr2008}. While it is not guaranteed that a pathology-proven cancer could have been determined with appropriate follow-up, it is likely that cancer existed at the time of acquisition for the majority of these exams. 

\begin{figure}[t]
	\begin{center}
		\includegraphics[width=1.0\textwidth]{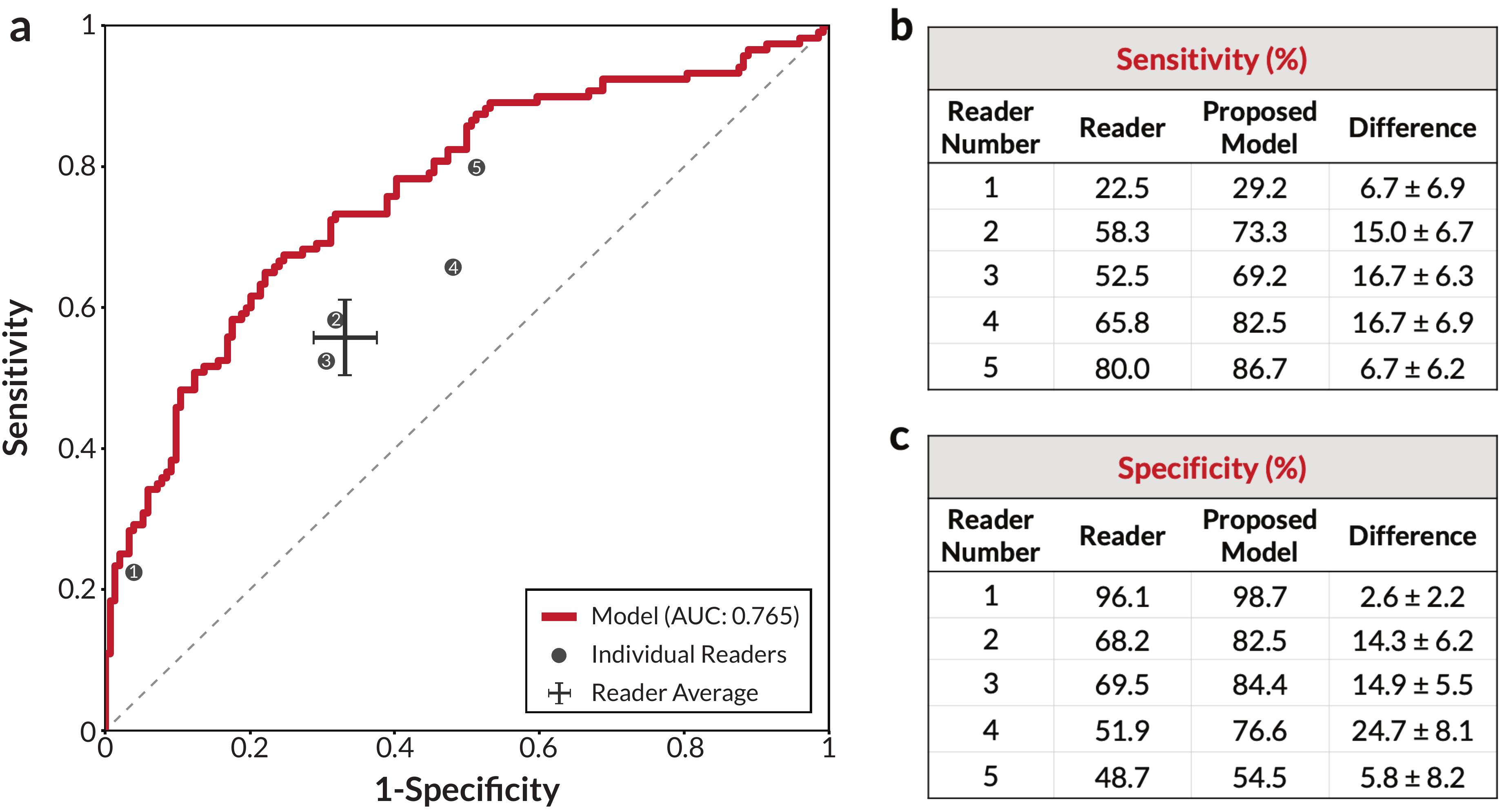}
	\end{center}
	\caption{\textbf{Reader study results -- pre-index cancer exams.} \textbf{a)} 
	The proposed deep learning model outperformed all five radiologists on the early detection task. 
	The dataset consisted of 120 pre-index cancer exams - which are defined as mammograms clinically interpreted as negative 12-24 months prior to the index exam in which cancer was found - and 154 confirmed negatives.
	Each data point represents a single reader, and the ROC curve represents the performance of the deep learning model. The cross corresponds to the mean radiologist performance with the lengths of the cross indicating 95\% confidence intervals for mean sensitivity and specificity. 
	\textbf{b)} Sensitivity of each reader and the corresponding sensitivity of the proposed model at a specificity chosen to match each reader. 
	\textbf{c)} Specificity of each reader and the corresponding specificity of the proposed model at a sensitivity chosen to match each reader.
	For \textbf{b)} and \textbf{c)}, the standard deviation of the model minus reader difference was calculated using 10K bootstrap samples.} 
	\label{gopal_pre-index}
\end{figure}

\indent As indicated in Figure~\ref{gopal_pre-index}, the deep learning model outperformed all five readers in the early detection, pre-index paradigm as well. The absolute performances of the readers and the deep learning model were lower on the pre-index cancer exams than on the index cancer exams, as expected given the difficulty of these cases. Nonetheless, the deep learning model still demonstrated an absolute increase in sensitivity of 17.5\% (95\% CI: 6.0-26.2\%; $p=0.001$) at the average reader specificity, and an absolute increase in specificity of 16.2\% (95\% CI: 7.3-24.6\%; $p=0.001$) at the average reader sensitivity. At a specificity of 90\%~\cite{LehmanAraoSpragueEtAl2017}, the model would have flagged 45.8\% (95\% CI: 28.8-57.1\%) of the pre-index cancer cases for additional workup.

\indent To further test the generalizability of our model, we assessed its performance on a DM dataset collected at an urban Chinese hospital. The model was evaluated locally at the hospital, with data never leaving the site. Testing generalization to this dataset is particularly meaningful given the low screening rates in China~\cite{fan2014breast} and the known (and potentially unknown) biological differences found in mammograms between Western and Asian populations. For instance, there is a greater proportion of dense breasts in Asian populations~\cite{bae2016breast}, which can increase the difficulty of mammogram interpretation. 

\indent Our deep learning model - which was trained primarily on Western populations - generalizes well to a Chinese population, achieving an AUC of 0.971 $\pm$ 0.005 (on the equivalent of ``index'' cancer exams; Table~\ref{china_dbt_table}). Even when adjusting for tumor size to approximately match the statistics expected in an American population, the model achieved 0.956 $\pm$ 0.020 AUC (see Table~\ref{china_dbt_table} and Methods). Table~\ref{china_dbt_table} additionally contains the results of the model when evaluated on held-out testing splits of OMI-DB and data from Site A, both also illustrating high levels of performance (see Extended Data Table~\ref{big-table-full} for further analysis).

\begin{table}[h]
	\begin{center}
		\includegraphics[width=0.7\textwidth]{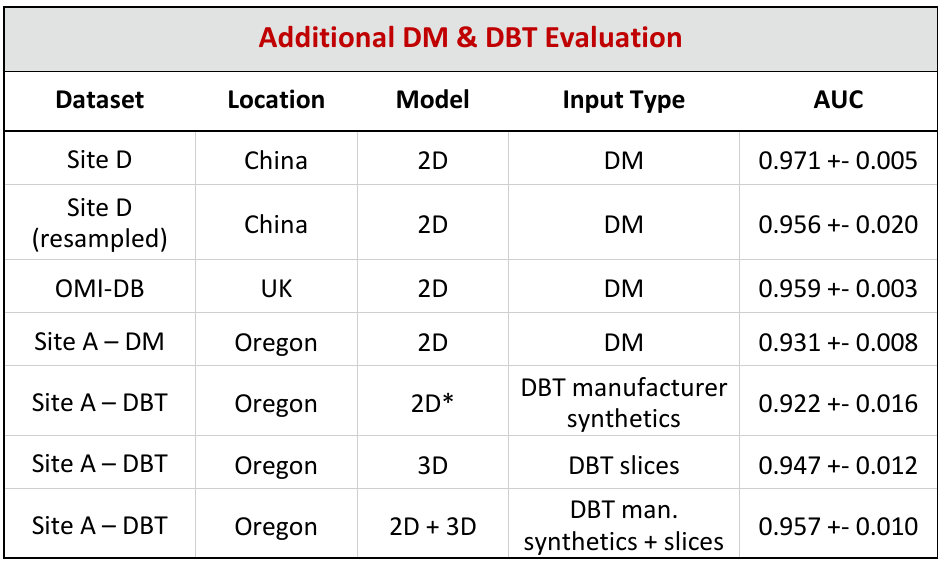}
	\end{center}
	\caption{\textbf{Summary of additional model evaluation.} All results correspond to using the ``index'' exam for cancer cases; ``pre-index'' results where possible and additional analysis is included in Extended Data Table~\ref{big-table-full}.
	Rows 1 \& 2: The 2D deep learning model, trained primarily on Western populations, performs well on a dataset collected at a Chinese hospital. Given the low prevalence of screening mammography in China, the Chinese data consists of diagnostic mammograms (i.e. mammograms of symptomatic patients). Nevertheless, even when adjusting for tumor size using bootstrap resampling to approximate the distribution of tumor sizes expected in an American screening population~\cite{nmd}, the model still achieves high performance (Row 2). Rows 3-4: Performance of the 2D deep learning model on held-out test sets of the OMI-DB and Site A datasets. Rows 5-7: Performance on DBT data. Row 5 contains results of the 2D model fine-tuned on the manufacturer-generated synthetic 2D images, which are created to augment/substitute DM images in a DBT study.
	Row 6 contains the results of the weakly-supervised 3D model, illustrating strong performance when evaluated on the DBT slices. 
	Row 7: Combining predictions across the final 3D model and 2D models results in even higher performance.}
    \label{china_dbt_table}
\end{table}

\indent Finally, our DBT approach performs well when evaluated at a site not used for DBT model training, as summarized in Table~\ref{china_dbt_table}. Our method, which generates an optimized image from the DBT slices and then classifies this image, achieves an AUC of 0.947 $\pm$ 0.012 (with index cancer exams). When scoring the DBT volume as the maximum bounding box score over all of the slices, the strongly-supervised 2D model used to create the optimized images exhibits an AUC of 0.865 $\pm$ 0.020. Thus, fine-tuning this model on the optimized DBT images significantly improves performance. Many DBT machines automatically generate synthetic 2D images from DBT volumes, and these synthetic images are often used as a supplement to or replacement for DM images in a DBT study; when fine-tuning our strongly-supervised 2D model on the manufacturer-generated synthetic images from the training set, the resulting model achieves 0.922 $\pm$ 0.016 AUC on the test set. Averaging predictions across the manufacturer-generated synthetic 2D images and our optimized 2D images results in an even higher overall performance of 0.957 $\pm$ 0.010.

\indent In summary, we have developed a deep learning approach that achieves state-of-the-art performance in classification of breast cancer in screening mammograms. By relying on a series of increasingly difficult training stages, we effectively leverage both strongly-labeled and weakly-labeled data from a wide array of data sources. The resulting model outperformed five-out-of-five full-time breast imaging specialists in a reader study involving data from a site not used for model training. This performance differential occurred on both the exams in which cancers were found clinically and the prior exams of these cancer cases which were deemed clinically-negative at the time of interpretation. 
The high performance of our model on data from an urban Chinese hospital further indicates its ability to generalize well across populations.
Lastly, our novel training approach for DBT - which utilizes transfer learning and a novel region of interest aggregation strategy - also results in high performance, even when using only weakly-labeled data. Altogether, these results show great promise towards earlier cancer detection and improved access to life-saving screening mammography using deep learning.

\subsubsection*{Acknowledgments}
This work was partly supported by the National Cancer Institute (NIH SBIR 1R44CA240022) and the National Science Foundation (NSF SBIR 1938387). We would additionally like to thank Dan Kopans, Etta Pisano, Polina Golland, Christoph Lee, Thomas Witzel, and Jacqui Holt for guidance and fruitful discussions.

\clearpage

\section*{Methods}
\vspace{-5pt}
\subsection*{Dataset Descriptions}\vspace{-6pt}
Details of all utilized datasets are provided below. All non-public datasets were collected under IRB approval and were de-identified prior to model training/testing. Within each data source, cross-validation splits were created at the patient level, meaning that exams from a given patient are all in the same cross-validation split.
Rules for label assignment and case selection for training data varied slightly across datasets given variability in collection time periods and available metadata. However, the definitions of testing sets and label criteria were standardized across datasets unless otherwise stated. In the main text, the following definitions were used in assigning labels: Index cancer - Screening exam obtained within three months preceding cancer diagnosis; Pre-index cancer - Screening exam interpreted as Breast Imaging Reporting And Data System (BIRADS~\cite{Sickles2013}) category 1 or 2 and obtained 12-24 months prior to index exam; Negative - Screening exam interpreted as BIRADS 1 or 2 and followed by an additional BIRADS 1 or 2 interpretation at the next screening exam 12-24 months later. In Extended Data Table~\ref{big-table-full}, we include additional results using a 12 month time window for defining an index cancer exam, as well as including pathology-proven benign cases.

\paragraph{Digital Database of Screening Mammography (DDSM)}
DDSM is a public database of scanned film mammography studies from the US containing normal, benign, and malignant cases with verified pathology information~\cite{Heath2000}. The dataset includes radiologist-drawn segmentation maps for every detected lesion. We split the data into 90\%/10\% training/model selection splits, resulting in 732 cancer, 743 benign, and 807 normal studies for training.

\paragraph{OPTIMAM Mammography Image Database (OMI-DB)}
The OMI-DB is a publicly available dataset from the UK, containing screening and diagnostic digital mammograms primarily obtained using Hologic and GE equipment~\cite{omidb}. We split the data into 60\%/20\%/20\% training/model selection/testing splits. We use the following definitions of cancer status for training: Cancer - Case acquired within two years of pathology-proven malignancy; Benign - Case acquired within one year of a benign biopsy from a patient who had no record of cancer in the database; Negative - A normal exam (BIRADS 1 \& 2) from a patient who had no record of cancer in the database. These definitions result in 5,233 cancers (2,332 with bounding boxes), 1,276 benigns (296 with bounding boxes), and 16,887 negatives for training.
The test results in Table~\ref{china_dbt_table} are calculated using 1217 cancer cases and 3000 negatives. Given the relatively low proportion of patients that have more than one screening exam in OMI-DB, we do not require a subsequent negative screening exam for the negative cases in the test set; instead we sample from the set of all negative exams from patients who have no record of cancer in the database.

\paragraph{Site A}
Site A consists of a community hospital in Oregon. The dataset from Site A primarily consists of screening mammograms, with DM data from 2010-2015 mostly collected from GE equipment, and DBT data from 2016-2017 collected from Hologic equipment. We split the DM data into 40\%/20\%/40\% training/model selection/testing splits. We use the DBT data solely for testing, given its high proportion of screening exams compared to the other utilized DBT datasets. Ground truth cancer status for both modalities was obtained using a local cancer registry. A report also accompanied each study and contained BIRADS information. For the DBT data, a list of benigns was also additionally provided by the hospital. For 2D data training, we use the following definitions of cancer status: Cancer - Screening case acquired within 15 months of pathology-proven malignancy; Negative - Screening mammogram interpreted as BIRADS 1 or 2, with no record of non-screening procedures or non-normal interpretations for the patient for 18 months prior to and following the exam. These definitions result in 466 cancers and 48,248 negatives for training. The DM testing results in Table~\ref{china_dbt_table} are calculated using 249 cancers and 1,990 negatives.
For DBT, the test results in Table~\ref{china_dbt_table} include 78 cancer cases and 519 negatives.

\paragraph{Site B}
Site B consists of an inpatient medical center in Rhode Island. The data from this site contains DBT mammograms, with a mix of screening and diagnostic exams collected retrospectively between 2016-2017. Cancer status, benign results, and BIRADS were determined using a local database. We split the dataset into 80\%/20\% training/model selection splits. We use the following definitions of cancer status for training: Cancer - Case acquired within 15 months of pathology-proven malignancy; Benign - Case with subsequent benign biopsy results from a patient with no record of cancer; Negative - Case interpreted as BIRADS 1 or 2. These definitions result in 13,767 negatives, 379 benigns, and 263 cancers for training. We note that the manufacturer-generated synthetic 2D images were also included in the weakly supervised training for the final 2D model.

\paragraph{Site C}
Site C consists of a health system in Oregon separate from the one in Site A. From Site C, DBT cases were retrospectively collected between 2017-2018. The data consists of a mix of screening and diagnostic cases, mostly acquired using Hologic equipment. We split the data into 70\%/30\% training/model selection splits. A local registry was used to determine cancer status. The following definitions were used for training: Cancer - Case acquired within 15 months of cancer status, as determined by entry in a regional cancer registry; Negative - Case from a patient with no entry ever in the regional cancer registry. These definitions result in 613 cancers and 8,355 negatives for training. Given the geographic proximity between Site C and Site A, we exclude a small number of patients that overlap in both sets when performing testing on Site A. We note that the manufacturer-generated synthetic 2D images were also included in the weakly supervised training for the final 2D model.

\paragraph{Site D}
Site D consists of a dataset from an urban hospital in China collected retrospectively from a contiguous period between 2012-2017. Used solely for model testing, the Site D dataset consists of 1,000 negatives (BIRADS 1 or 2 interpretation), 533 pathology-proven cancers, and 100 pathology-proven benigns. Given low screening rates in China, the data came from diagnostic exams (i.e. exams where the patient presented with symptoms), so the distribution of tumor sizes from the cancer cases contained more large tumors (64\% larger than 2 cm) than would be expected in a typical United States screening population. To better compare to a US screening population, results on Site D are also calculated using a bootstrap resampling method to approximately match the distribution of tumor sizes from a US population according to the National Radiology Data Registry~\cite{nmd}. Using this approach, a mean AUC is computed over 5K bootstrapped populations.

\subsection*{Model Development and Training}\vspace{-6pt}
\indent The first stage of model training consisted of patch-based classification~\cite{Lotter2017}. Patches of size 275x275 pixels were created from the DDSM and Optimam datasets after the original images were resized to a height of 1750 pixels. Data augmentation was also used when creating the patches, including random rotations and image resizing by up to 20\%. The patch-based training stage itself consisted of two training sequences. First, starting from ImageNet~\cite{imagenet} pre-trained weights, a ResNet-50 model~\cite{He2016} was trained for five-way classification of lesion type: mass, calcifications, focal asymmetry, architectural distortion, or no lesion. The model was trained for 62,500 batches with a batch size of 16, sampling equally from all lesion types. The Adam optimizer~\cite{adam} was used with a learning rate of $1\mathrm{e}{-5}$. Next, the patch-level model was trained for three-way classification, using labels of normal, benign, or malignant, again sampling equally from all categories. The same training parameters were also used for this stage of patch-level training.

\indent After patch-level training, the ResNet-50 weights were used to initialize the backbone of a RetinaNet model~\cite{retinanet} for the second stage of training: strongly-supervised, image-level training. Image pre-processing consisted of resizing to a height of 1750 pixels (maintaining the original aspect ratio), cropping out the background, and normalizing pixel values to a range of [$-127.5$, $127.5$]. Data augmentation during training included random resizing of up to 15\% and random vertical mirroring.

\indent The strongly-supervised, image-level training stage involved training with bounding boxes. Training examples were sampled from OMI-DB and DDSM based on the relative proportion of malignant images in each dataset. Within each dataset, malignant and non-malignant examples were sampled with equal probability, using batch sizes of 1. Three-way bounding box classification was performed using labels of normal, benign, or malignant. For this training stage, only strongly-labeled examples were included for malignant images. The RetinaNet model was trained for 100K iterations, with a batch size of 1. The Adam optimizer~\cite{adam} was used, with a learning rate of $1\mathrm{e}{-5}$ and gradient norm clipping with a value of 0.001. Default hyperparameters were used in the RetinaNet loss, except for a weight of 0.5 that was given to the regression loss and a weight of 1.0 that was given to the classification loss. 

\indent For the weakly-supervised training stage, binary cancer/no-cancer classification was performed with a binary cross entropy loss. The same image input processing steps were used as in the strongly-supervised training stage. For 2D, training consisted of 300K iterations using the Adam optimizer~\cite{adam}, starting with a learning rate of $2.5\mathrm{e}{-6}$, which was decreased by a factor of 4 every 100K iterations. Final model weights were chosen by monitoring AUC performance on the validation set every 4K iterations. 
After the weakly-supervised training stage, we note that it is also feasible in the RetinaNet framework to fine-tune the regression head of the network again on strongly-supervised data while freezing the backbone and classification head.

\indent For DBT weakly-supervised training, the optimized images were created using the model resulting from 2D strongly-supervised training as described above, after an additional 50K training iterations with a learning rate of $2.5\mathrm{e}{-6}$. To create the optimized images, the 2D model was evaluated on every slice in a DBT stack except for the first and last 10\% of slices (which are frequently noisy). A minimal bounding box score threshold was set at a level that achieved 99\% sensitivity on the OMI-DB validation set. An intersection-over-union (IOU) threshold of 0.2 was used for NMS filtering. Any “empty” pixels in the optimized images were infilled with the corresponding pixels from the center slice of the DBT stack. The input processing parameters used for 2D images were reused for DBT slices and optimized images. Training on the optimized images was conducted similarly to 2D weakly-supervised training, except that the model was trained for 100K iterations.

\indent Final 2D and 3D models were created by aggregating three similarly-trained models. The final prediction score for a given image was calculated by averaging across the three models’ predictions for both horizontal orientations of the image (i.e. an average over six scores). Each breast was assigned a score by taking the average score of all its views. Each study was assigned a score by taking the greater of its two breast-level scores.

\indent All models were developed and evaluated using the Keras library~\cite{keras} with a Tensorflow backend~\cite{tensorflow}.

\subsection*{Reader Study}
\paragraph{Data Collection}
A total of 405 screening DM exams were collected retrospectively from a single health system in Massachusetts under an IRB-approved protocol compliant with the Health Insurance Portability and Accountability Act. The exams were acquired between July 2011 and June 2014. 154 of the exams were negative, receiving an interpretation of BIRADS 1 or BIRADS 2, and all negative exams were followed by at least one additional subsequent negative screening exam. 131 of the exams came from women whose mammograms were initially interpreted as suspicious and who were diagnosed with cancer within three months of the screening date by pathological examination of their biopsies. These exams were termed ``index'' cancer exams. Additionally, previous screening exams conducted 12-24 months prior to the “index” exams were collected for 120 out of the 131 women, and were labelled ``pre-index''. These ``pre-index'' mammograms were interpreted as BIRADS 1 or BIRADS 2 at the time of initial clinical examination. In total, 154 non-cancer, 131 index  cancer, and 120 pre-index cancer mammograms were collected from 285 women. The non-cancer cases were chosen to have a similar distribution in patient age and breast density compared to the cancer cases. 

\paragraph{Reader Selection}
Five board-certified and MQSA-qualified radiologists were recruited as readers. All readers were fellowship trained in breast imaging and had practiced for an average of 5.6 years post-fellowship (range 2-15 years). The readers read an average of 6,969 mammograms over the year preceding the reader study (range of 2,921 - 9,260), 60\% of which were DM and 40\% of which were DBT.

\paragraph{Study Design}
The reader study was conducted in two sessions. During the first session, radiologists read the 131 index cancer exams and 76 of the negative exams. During the second session, radiologists read the 120 pre-index exams and the remaining 78 negative exams. There was a washout period of at least 4 weeks in between the two sessions for each reader. The readers were instructed to give a forced BIRADS score for each case (1 - 5). BIRADS 1 and 2 were considered no recall, and BIRADS 3, 4, and 5 were considered recall~\cite{LehmanAraoSpragueEtAl2017}. Radiologists did not have any information about the patients (such as previous medical history, radiology reports, and other patient records), and were informed that the study dataset is enriched with cancer mammograms relative to the standard prevalence observed in screening; however, they were not informed about the proportion of case types. All radiologists viewed and interpreted the studies on dedicated mammography workstations in an environment similar to their clinical practice.

\subsection*{Statistical Analysis}
The confidence intervals for the model AUCs and average readers’ sensitivity and specificity were computed based on percentiles of a bootstrap method with 10,000 random resamples. The p-value for comparing the model’s sensitivity and specificity with the average reader sensitivity and specificity was computed by taking the proportion of times the difference between the model and readers was less than 0 across the resampled data. The p-value for comparing AUCs between two models was computed using the DeLong method~\cite{DeLong1988}.

\clearpage

\section*{Extended Data}

\renewcommand{\figurename}{\textbf{Extended Data Figure}}
\setcounter{figure}{0}

\renewcommand{\tablename}{\textbf{Extended Data Table}}
\setcounter{table}{0}

\begin{figure}[h]
	\begin{center}
		\includegraphics[width=1.0\textwidth]{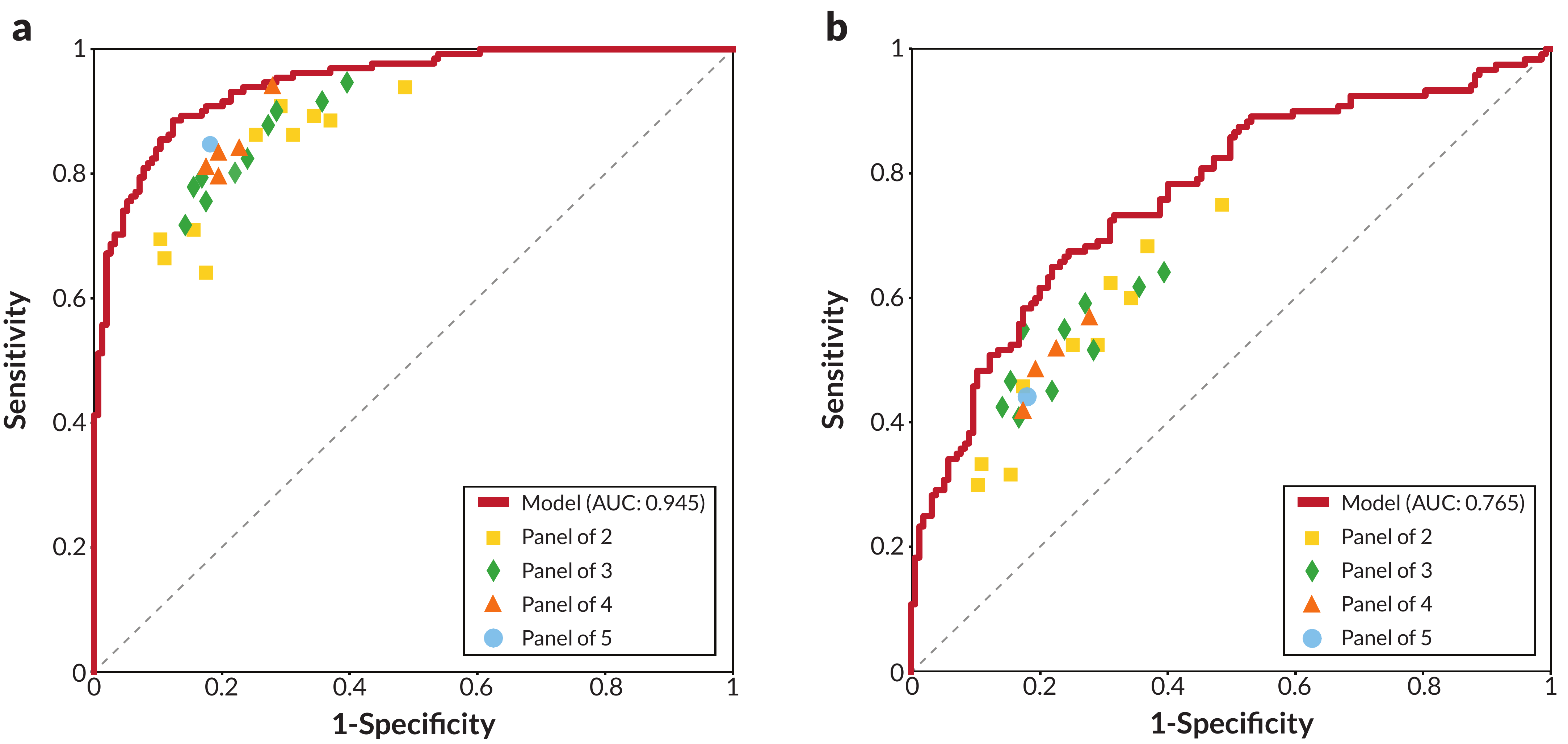}
	\end{center}
	\caption{\textbf{Results of model compared to synthesized panel of readers.}
	Comparison of model ROC curves to every combination of 2, 3, 4, and 5 readers.
	Readers were combined by averaging BIRADS scores, with sensitivity and specificity calculated using a threshold of 3.
	On both the \textbf{a)} index cancer exams and \textbf{b)} pre-index cancer exams, the model outperformed every combination of readers, as indicated by each combination falling below the model's respective ROC curve.
	} 
	\label{panel_figure}
\end{figure}

\clearpage

\begin{table}[h]
\caption{\textbf{Comparison to recent work -- index cancer exams.} Below we compare the performance of our proposed model to other recently proposed models on the set of index cancer exams and confirmed negatives from our reader study. We compare to the models of Yala et al., 2019~\cite{Yala2019}, Wu et al., 2019~\cite{Wu2019}, and to the model that achieved the highest final score on the Digital Mammography DREAM Challenge~\cite{noauthor_digital_nodate, Wu2019ValidationRates}.} 
\label{model_comparison_table}
\begin{subtable}{\textwidth}
    \caption{\textbf{AUC comparison.} The proposed model had a significantly higher AUC than the other models under comparison. P-values for AUC differences were calculated using the DeLong method. Confidence intervals were calculated using 10K bootstrap samples.}
    \label{Tab:auc_models_index}
    \centering
    \begin{tabular}{llll}
        \toprule
        Model & AUC (95\% CI)        & Delta to & p-value    \\
                          &  &           Proposed            &                  \\
        \midrule
        \textbf{Proposed Model} & \textbf{0.945 (0.919, 0.968)} &        &            \\
        DREAM~\cite{Wu2019ValidationRates} & 0.895 (0.855, 0.931) & 0.050 & 0.0006     \\
        Yala et al., 2019~\cite{Yala2019}   & 0.858 (0.812, 0.900) & 0.087 & 1.14E-06    \\
        Wu et al., 2019~\cite{Wu2019}    & 0.841 (0.792, 0.887) & 0.104  & 1.97E-06   \\
        \bottomrule
    \end{tabular}
\end{subtable}

\begin{subtable}{\textwidth}
    \caption{\textbf{Sensitivity of models compared to readers.} Sensitivity was obtained at the point on the ROC curve corresponding to the average reader specificity. Delta values show the difference between model sensitivity and average reader sensitivity and the p-values correspond to this difference. Confidence intervals and p-values were calculated using 10K bootstrap samples.}
    \label{Tab:sensitivity_model_v_reader_index}
    \centering
    \begin{tabular}{llll}
        \toprule
        Model             & Sensitivity (95\% CI)& Delta to Readers                & p-value          \\
                          & @ Reader Specificity &                   &                  \\
        \midrule
        \textbf{Proposed Model}             & \textbf{96.2\% (91.7\%, 99.2\%)} & \textbf{14.2\% (9.2\%,  18.5\%)}  & \textbf{\textless 0.0001} \\
        DREAM~\cite{Wu2019ValidationRates}            & 91.6\% (84.3\%, 95.9\%) & 9.6\% (2.2\%, 14.8\%)  & 0.003           \\
        Yala et al., 2019~\cite{Yala2019} & 86.3\% (75.7\%, 92.9\%) & 4.3\% (-6.8\%, 11.7\%) & 0.205           \\
        Wu et al., 2019~\cite{Wu2019}              & 84.0\% (74.4\%, 90.2) & 2.0\% (-7.8\%, 8.9\%) & 0.372           \\
        \bottomrule
    \end{tabular}
\end{subtable}

\begin{subtable}{\textwidth}
    \caption{\textbf{Specificity of models compared to readers.} Specificity was obtained at the point on the ROC curve corresponding to the average reader sensitivity. Delta values show the difference between model specificity and average reader specificity and the p-values correspond to this difference. Confidence intervals and p-values were calculated using 10K bootstrap samples.}
    \label{Tab:specificity_model_v_reader_index}
    \centering
    \begin{tabular}{llll}
        \toprule
        Model & Specificity (95\% CI)& Delta to Readers                 & p-value          \\
              & @ Reader Sensitivity &                       &                  \\
        \midrule
        \textbf{Proposed Model} & \textbf{90.9\% (84.9\%, 96.1\%)}  & \textbf{24.0\% (17.4\%, 30.4\%)}  & \textbf{\textless 0.0001} \\
        DREAM~\cite{Wu2019ValidationRates} & 85.7\% (73.4\%, 92.9\%) & 18.8\% (6.4\%, 26.8\%)  & 0.003            \\
        Yala et al., 2019~\cite{Yala2019}   & 71.4\% (59.6\%, 84.5\%)  & 4.5\% (-5.7\%, 13.0\%) & 0.244           \\
        Wu et al., 2019~\cite{Wu2019}   & 68.2\% (48.5\%, 84.0\%)  & 1.3\% (-18.5\%, 17.5\%) & 0.369           \\
        \bottomrule
    \end{tabular}
\end{subtable}
\end{table}

\clearpage

\begin{table}[h]
\caption{\textbf{Comparison to recent work -- pre-index cancer exams.} Below we compare the performance of our proposed model to other recently proposed models on the set of pre-index cancer exams and confirmed negatives from our reader study. We compare to the models of Yala et al., 2019~\cite{Yala2019}, Wu et al., 2019~\cite{Wu2019}, and to the model that achieved the highest final score on the Digital Mammography DREAM Challenge~\cite{noauthor_digital_nodate, Wu2019ValidationRates}.}
\begin{subtable}{\textwidth}
    \caption{\textbf{AUC comparison.} The proposed model had a significantly higher AUC than the other models under comparison. P-values for AUC differences were calculated using the DeLong method. Confidence intervals were calculated using 10K bootstrap samples.}
    \label{Tab:auc_models_pre_index}
    \centering
    \begin{tabular}{llll}
        \toprule
        Model & AUC (95\% CI)        & Delta to  & p-value \\
                   & &        Proposed               &                  \\
        \midrule
        \textbf{Proposed Model} & \textbf{0.765 (0.705, 0.820)} &        &         \\
        DREAM \cite{Wu2019ValidationRates} & 0.701 (0.639, 0.763) & 0.064  & 0.009 \\
        Yala et al., 2019 \cite{Yala2019}   & 0.702 (0.638, 0.762) & 0.063  & 0.008 \\
        Wu et al., 2019 \cite{Wu2019}   & 0.701 (0.637, 0.762) & 0.064  & 0.040 \\
        \bottomrule
    \end{tabular}
\end{subtable}

\begin{subtable}{1.0\textwidth}
    \caption{\textbf{Sensitivity of models compared to readers.} Sensitivity was obtained at the point on the ROC curve corresponding to the average reader specificity. Delta values show the difference between model sensitivity and average reader sensitivity and the p-values correspond to this difference. Confidence intervals and p-values were calculated using 10K bootstrap samples.}
    \label{Tab:sensitivity_model_v_reader_pre_index}
    \centering
    \begin{tabular}{llll}
        \toprule
        Model & Sensitivity (95\% CI)& Delta to Readers               & p-value          \\
              & @ Reader Specificity &                       &                  \\
        \midrule
        \textbf{Proposed Model} & \textbf{73.3\% (62.0\%, 81.3\%)} & \textbf{17.5\% (6.0\%, 26.2\%)}  & \textbf{0.0009} \\
        DREAM \cite{Wu2019ValidationRates} & 60.8\% (49.2\%, 76.5\%) & 5.0\% (-5.7\%, 20.5\%) & 0.182 \\
        Yala et al., 2019 \cite{Yala2019}   & 59.2\% (49.1\%, 68.9\%) & 3.4\% (-7.5\%, 17.8\%) & 0.204 \\
        Wu et al., 2019 \cite{Wu2019}   & 66.7\% (51.9\%, 75.4\%) & 10.9\% (-3.8\%, 19.5\%) & 0.072 \\
        \bottomrule
    \end{tabular}
\end{subtable}

\begin{subtable}{1.0\textwidth}
    \caption{\textbf{Specificity of models compared to readers.} Specificity was obtained at the point on the ROC curve corresponding to the average reader sensitivity. Delta values show the difference between model specificity and average reader specificity and the p-values correspond to this difference. Confidence intervals and p-values were calculated using 10K bootstrap samples.}
    \label{Tab:specificity_model_v_reader_pre_index}
    \centering
    \begin{tabular}{llll}
        \toprule
        Model & Specificity (95\% CI)  & Delta to Readers                 & p-value\\
              & @ Reader Sensitivity   &                        &        \\
        \midrule
        \textbf{Proposed Model} & \textbf{83.1\% (74.5\%, 91.0\%)}   & \textbf{16.2\% (7.3\%, 24.6\%)}   & \textbf{0.0008} \\
        DREAM \cite{Wu2019ValidationRates} & 72.1\% (61.5\%, 82.5\%)   & 5.2\% (-5.4\%, 15.7\%)  & 0.179 \\
        Yala et al., 2019 \cite{Yala2019}   & 73.4\% (56.2\%, 86.9\%)   & 6.5\% (-10.5\%, 20.3\%)  & 0.252 \\
        Wu et al., 2019 \cite{Wu2019}   & 73.4\% (64.9\%, 85.4\%)   & 6.5\% (-2.5\%, 19.4\%)  & 0.070 \\
        \bottomrule
    \end{tabular}
\end{subtable}
\end{table}

\clearpage

\begin{table}[h]
\caption{\textbf{Performance of proposed models under different case compositions.}
In the main text, we chose case compositions and definitions to match those of the reader study, specifically index cancer exams were mammograms acquired within 3 months preceding a cancer diagnosis and non-cancers were negative mammograms (BIRADS 1 or 2) that were followed by another negative mammogram the subsequent screen. Here, we additionally consider a 12 month definition of index cancer exams, i.e., mammograms acquired within 0-12 months preceding a cancer diagnosis. The number of cancer cases added for the 12 month definition compared to the 3 month definition is 39, 46, and 7 for OMI-DB, Site A - DM, and Site A - DBT, respectively.
A 12-24 month time window results in 68 cancer cases for OMI-DB and 232 cancer cases for Site A - DM.
For a given dataset, the negative cases are shared amongst all cancer time window calculations, with case counts detailed in the Methods.
In \textbf{b)}, we also include pathology-proven benigns amongst the non-cancers. These cases correspond to mammograms in which the patient was recalled and ultimately biopsied, but the biopsy was benign. We include a 10\% benign rate with respect to the total number of negatives for each calculation, corresponding to a typical recall rate in the United States~\cite{LehmanAraoSpragueEtAl2017}. 
Dashes indicate calculations that are not possible given the data and information available for each site.
}
\label{big-table-full}
\begin{subtable}{\textwidth}
\centering
\caption{\textbf{AUC by cancer time window -- no benigns.}}

\begin{tabular}{llll}
\toprule
Dataset                              & 0-3 months     & 0-12 months   & 12-24 months  \\
\midrule
OMI-DB - UK                                 & 0.959 $\pm$ 0.003  & 0.954 $\pm$ 0.003 & 0.729 $\pm$ 0.029   \\
Site D - China (Original)                     & 0.971 $\pm$ 0.005   & --               & --               \\
Site D - China (Resampled)                    & 0.956 $\pm$ 0.020   & --               & --               \\
Site A - Oregon (DM)                        & 0.931 $\pm$ 0.008  & 0.907 $\pm$ 0.010 & 0.763 $\pm$ 0.017 \\
\midrule
Site A - Oregon (DBT) & & & \\
\hspace{3mm}Slices, no fine-tuning      & 0.865 $\pm$ 0.020   & 0.859 $\pm$ 0.021  & --               \\
\hspace{3mm}Manufacturer synthetics, no fine-tuning     & 0.893 $\pm$ 0.022   & 0.885 $\pm$ 0.022  & --               \\
\hspace{3mm}Manufacturer synthetics, with fine-tuning        & 0.922 $\pm$ 0.016  & 0.917 $\pm$ 0.016 & --               \\
\hspace{3mm}Slices, final model       & 0.947 $\pm$ 0.012 & 0.936 $\pm$ 0.014 & --               \\
\hspace{3mm}Slices \& man. synthetics, final models & 0.957 $\pm$ 0.010  & 0.945 $\pm$ 0.013  & --               \\
\bottomrule
\end{tabular}

\label{big-no-benigns}
\end{subtable}

\begin{subtable}{\textwidth}
\centering
\caption{\textbf{AUC by cancer time window -- with benigns.}} 
\begin{tabular}{llll}
\toprule
Dataset                              & 0-3 months     & 0-12 months   & 12-24 months  \\
\midrule
OMI-DB - UK                                   & 0.948 $\pm$ 0.004       & 0.943 $\pm$ 0.004  & 0.700 $\pm$ 0.030     \\
Site D - China (All data)                     & 0.966 $\pm$ 0.006        & --                & --                \\
Site D - China (Resampled)                   & 0.949 $\pm$ 0.020        & --                & --                \\
Site A - Oregon (DM)                             & --                     & --                & --                \\
\midrule
Site A - Oregon (DBT) & & & \\
\hspace{3mm}Slices, no fine-tuning     & 0.863 $\pm$ 0.020        & 0.856 $\pm$ 0.021   & --                \\
\hspace{3mm}Manufacturer synthetics, no fine-tuning        & 0.887 $\pm$ 0.023      & 0.879 $\pm$ 0.023 & --                \\
\hspace{3mm}Manufacturer synthetics, with fine-tuning     & 0.915 $\pm$ 0.017        & 0.910 $\pm$ 0.017   & --                \\
\hspace{3mm}Slices, final model        & 0.941 $\pm$ 0.012      & 0.930 $\pm$ 0.014   & --                \\
\hspace{3mm}Slices \& man. synthetics, final models & 0.950 $\pm$ 0.010          & 0.938 $\pm$ 0.013 & --                \\
\bottomrule
\end{tabular}

\label{big-with-benigns}
\end{subtable}
\end{table}

\clearpage

\bibliography{references_dl.bib}

\begin{thebibliography}{10}\itemsep=-1pt

\bibitem{noauthor_digital_nodate}
{Digital Mammography DREAM Challenge – Sage Bionetworks}.
\newblock
  \url{http://sagebionetworks.org/research-projects/digital-mammography-dream-challenge/}.

\bibitem{omidb}
Optimam mammography imaging.
\newblock \url{https://medphys.royalsurrey.nhs.uk/omidb/}.

\bibitem{tensorflow}
M.~Abadi, A.~Agarwal, P.~Barham, E.~Brevdo, Z.~Chen, C.~Citro, G.~S. Corrado,
  A.~Davis, J.~Dean, M.~Devin, S.~Ghemawat, I.~Goodfellow, A.~Harp, G.~Irving,
  M.~Isard, Y.~Jia, R.~Jozefowicz, L.~Kaiser, M.~Kudlur, J.~Levenberg,
  D.~Man\'{e}, R.~Monga, S.~Moore, D.~Murray, C.~Olah, M.~Schuster, J.~Shlens,
  B.~Steiner, I.~Sutskever, K.~Talwar, P.~Tucker, V.~Vanhoucke, V.~Vasudevan,
  F.~Vi\'{e}gas, O.~Vinyals, P.~Warden, M.~Wattenberg, M.~Wicke, Y.~Yu, and
  X.~Zheng.
\newblock {TensorFlow}: Large-scale machine learning on heterogeneous systems,
  2015.
\newblock Software available from tensorflow.org.

\bibitem{nmd}
{American College of Radiology}.
\newblock {NMD Registry: National Mammography Database}.
\newblock \url{https://nrdr.acr.org/Portal/NMD/Main/page.aspx}, 2019.

\bibitem{curemetrix_FDA}
D.~Anderson.
\newblock {CureMetrix receives FDA Clearance for AI-based triage software for
  mammography}, 2019.

\bibitem{bae2016breast}
J.-M. Bae and E.~H. Kim.
\newblock Breast density and risk of breast cancer in asian women: a
  meta-analysis of observational studies.
\newblock {\em Journal of Preventive Medicine and Public Health}, 49(6):367,
  2016.

\bibitem{Berry2005}
D.~A. Berry, K.~A. Cronin, S.~K. Plevritis, D.~G. Fryback, L.~Clarke, M.~Zelen,
  J.~S. Mandelblatt, A.~Y. Yakovlev, J.~D.~F. Habbema, and E.~J. Feuer.
\newblock {Effect of Screening and Adjuvant Therapy on Mortality from Breast
  Cancer}.
\newblock {\em New England Journal of Medicine}, 353(17):1784--1792, 2005.

\bibitem{Bray2018}
F.~Bray, J.~Ferlay, I.~Soerjomataram, R.~L. Siegel, L.~A. Torre, and A.~Jemal.
\newblock {Global cancer statistics 2018: GLOBOCAN estimates of incidence and
  mortality worldwide for 36 cancers in 185 countries}.
\newblock {\em CA: A Cancer Journal for Clinicians}, 68(6):394--424, 2018.

\bibitem{pmlr-v81-buolamwini18a}
J.~Buolamwini and T.~Gebru.
\newblock Gender shades: Intersectional accuracy disparities in commercial
  gender classification.
\newblock In S.~A. Friedler and C.~Wilson, editors, {\em Proceedings of the 1st
  Conference on Fairness, Accountability and Transparency}, volume~81 of {\em
  Proceedings of Machine Learning Research}, pages 77--91, New York, NY, USA,
  23--24 Feb 2018. PMLR.

\bibitem{keras}
F.~Chollet et~al.
\newblock Keras.
\newblock \url{https://keras.io}, 2015.

\bibitem{Conant2019}
E.~F. Conant, A.~Y. Toledano, S.~Periaswamy, S.~V. Fotin, J.~Go, J.~E.
  Boatsman, and J.~W. Hoffmeister.
\newblock {Improving Accuracy and Efficiency with Concurrent Use of Artificial
  Intelligence for Digital Breast Tomosynthesis}.
\newblock {\em Radiology: Artificial Intelligence}, 2019.

\bibitem{DeLong1988}
E.~R. DeLong, D.~M. DeLong, and D.~L. Clarke-Pearson.
\newblock {Comparing the Areas under Two or More Correlated Receiver Operating
  Characteristic Curves: A Nonparametric Approach}.
\newblock {\em Biometrics}, 1988.

\bibitem{imagenet}
J.~Deng, W.~Dong, R.~Socher, L.-J. Li, K.~Li, and L.~Fei-Fei.
\newblock {ImageNet: A Large-Scale Hierarchical Image Database}.
\newblock In {\em CVPR09}, 2009.

\bibitem{fan2014breast}
L.~Fan, K.~Strasser-Weippl, J.-J. Li, J.~St~Louis, D.~M. Finkelstein, K.-D. Yu,
  W.-Q. Chen, Z.-M. Shao, and P.~E. Goss.
\newblock Breast cancer in china.
\newblock {\em The lancet oncology}, 15(7):e279--e289, 2014.

\bibitem{mqsa_national_statistics}
{FDA}.
\newblock {MQSA National Statistics}.
\newblock
  \url{https://www.fda.gov/radiation-emitting-products/mqsa-insights/mqsa-national-statistics},
  2019.

\bibitem{Fenton2007}
J.~J. Fenton, S.~H. Taplin, P.~A. Carney, L.~Abraham, E.~A. Sickles, C.~D'Orsi,
  E.~A. Berns, G.~Cutter, R.~E. Hendrick, W.~E. Barlow, and J.~G. Elmore.
\newblock {Influence of Computer-Aided Detection on Performance of Screening
  Mammography}.
\newblock {\em New England Journal of Medicine}, 356(14):1399--1409, 4 2007.

\bibitem{Freer_2001}
T.~W. Freer and M.~J. Ulissey.
\newblock Screening mammography with computer-aided detection: Prospective
  study of 12,860 patients in a community breast center.
\newblock {\em Radiology}, 220(3):781--786, 2001.

\bibitem{geras2017highresolution}
K.~J. Geras, S.~Wolfson, Y.~Shen, N.~Wu, S.~G. Kim, E.~Kim, L.~Heacock,
  U.~Parikh, L.~Moy, and K.~Cho.
\newblock High-resolution breast cancer screening with multi-view deep
  convolutional neural networks.
\newblock {\em arXiv}, 2017.

\bibitem{Hart1998}
D.~Hart, E.~Shochat, and Z.~Agur.
\newblock {The growth law of primary breast cancer as inferred from mammography
  screening trials data}.
\newblock {\em British Journal of Cancer}, 78(3):382--387, 1998.

\bibitem{He2016}
K.~He, X.~Zhang, S.~Ren, and J.~Sun.
\newblock {Deep Residual Learning for Image Recognition}.
\newblock In {\em The IEEE Conference on COmputer Vision and Pattern
  Recognition (CVPR)}, pages 770--778, 2016.

\bibitem{Heath2000}
M.~Heath, D.~Kopans, R.~Moore, and {Kegelmeyer Jr. P}.
\newblock {The digital database for screening mammography}.
\newblock In {\em The British Journal of Psychiatry}, pages 212--218.
  Proceedings of the 5th international workshop on digital mammography, 2000.

\bibitem{Henriksen2019}
E.~L. Henriksen, J.~F. Carlsen, I.~M. Vejborg, M.~B. Nielsen, and C.~A.
  Lauridsen.
\newblock {The efficacy of using computer-aided detection (CAD) for detection
  of breast cancer in mammography screening: a systematic review.}
\newblock {\em Acta radiologica (Stockholm, Sweden : 1987)}, 60(1):13--18, 1
  2019.

\bibitem{Hoff2011}
S.~R. Hoff, J.~H. Samset, A.~L. Abrahamsen, E.~Vigeland, O.~Klepp, and
  S.~Hofvind.
\newblock {Missed and True Interval and Screen-detected Breast Cancers in a
  Population Based Screening Program}.
\newblock {\em Academic Radiology}, 18(4):454--460, 2011.

\bibitem{Ikeda2003}
D.~M. Ikeda, R.~L. Birdwell, K.~F. O'Shaughnessy, R.~J. Brenner, and E.~A.
  Sickles.
\newblock {Analysis of 172 subtle findings on prior normal mammograms in women
  with breast cancer detected at follow-up screening}.
\newblock {\em Radiology}, 226(2):494--503, 2003.

\bibitem{kim_design_2019}
D.~W. Kim, H.~Y. Jang, K.~W. Kim, Y.~Shin, and S.~H. Park.
\newblock {Design Characteristics of Studies Reporting the Performance of
  Artificial Intelligence Algorithms for Diagnostic Analysis of Medical Images:
  Results from Recently Published Papers}.
\newblock {\em Korean Journal of Radiology}, 20(3):405--410, 3 2019.

\bibitem{adam}
D.~P. Kingma and J.~Ba.
\newblock {Adam: A Method for Stochastic Optimization}.
\newblock In {\em 3rd International Conference on Learning Representations,
  ICLR 2015 - Conference Track Proceedings}. International Conference on
  Learning Representations, ICLR, 2015.

\bibitem{Kooi2017}
T.~Kooi, G.~Litjens, B.~van Ginneken, A.~Gubern-M{\'{e}}rida, C.~I.
  S{\'{a}}nchez, R.~Mann, A.~den Heeten, and N.~Karssemeijer.
\newblock {Large scale deep learning for computer aided detection of
  mammographic lesions}.
\newblock {\em Medical Image Analysis}, 35:303--312, 2017.

\bibitem{Kopans_2014}
D.~B. Kopans.
\newblock {Digital Breast Tomosynthesis From Concept to Clinical Care}.
\newblock {\em American Journal of Roentgenology}, 2014.

\bibitem{LehmanAraoSpragueEtAl2017}
C.~D. Lehman, R.~F. Arao, B.~L. Sprague, J.~M. Lee, D.~S.~M. Buist,
  K.~Kerlikowske, L.~M. Henderson, T.~Onega, A.~N.~A. Tosteson, G.~H. Rauscher,
  and D.~L. Miglioretti.
\newblock {National Performance Benchmarks for Modern Screening Digital
  Mammography: Update from the Breast Cancer Surveillance Consortium}.
\newblock {\em Radiology}, 283(1):49--58, 2017.

\bibitem{Lehman2016}
C.~D. Lehman, R.~D. Wellman, D.~S. Buist, K.~Kerlikowske, A.~N. Tosteson, and
  D.~L. Miglioretti.
\newblock {Diagnostic accuracy of digital screening mammography with and
  without computer aided detection.}
\newblock {\em JAMA Internal Medicine}, 33(8):839--841, 2016.

\bibitem{retinanet}
T.~{Lin}, P.~{Goyal}, R.~{Girshick}, K.~{He}, and P.~{Dollár}.
\newblock Focal loss for dense object detection.
\newblock In {\em 2017 IEEE International Conference on Computer Vision
  (ICCV)}, pages 2999--3007, 2017.

\bibitem{Lotter2017}
W.~Lotter, G.~Sorensen, and D.~Cox.
\newblock {A Multi-Scale CNN and Curriculum Learning Strategy for Mammogram
  Classification}.
\newblock In {\em Deep Learning in Medical Image Analysis and Multimodal
  Learning for Clinical Decision Support}, 2017.

\bibitem{Mainiero2017}
M.~B. Mainiero, L.~Moy, P.~Baron, A.~D. Didwania, R.~M. DiFlorio, E.~D. Green,
  S.~L. Heller, A.~I. Holbrook, S.-J. Lee, A.~A. Lewin, A.~P. Lourenco, K.~J.
  Nance, B.~L. Niell, P.~J. Slanetz, A.~R. Stuckey, N.~S. Vincoff, S.~P.
  Weinstein, M.~M. Yepes, and M.~S. Newell.
\newblock {ACR Appropriateness Criteria{\textregistered} Breast Cancer
  Screening}.
\newblock {\em Journal of the American College of Radiology}, 14(11):S383 --
  S390, 2017.

\bibitem{Majid2003}
A.~S. Majid, E.~Shaw De~Paredes, R.~D. Doherty, N.~R. Sharma, and X.~Salvador.
\newblock {Missed Breast Carcinoma: Pitfalls and Pearls}.
\newblock {\em RadioGraphics}, 23:881--895, 2003.

\bibitem{NHSBSP2016}
{NHS Breast Screening Programme}.
\newblock {Clinical guidance for breast cancer screening assessment}.
\newblock NHSBSP publication no 49. 4th edn, 2016.

\bibitem{Oeffinger2015}
K.~C. Oeffinger, E.~T. Fontham, R.~Etzioni, A.~Herzig, J.~S. Michaelson, Y.~C.
  Shih, L.~C. Walter, T.~R. Church, C.~R. Flowers, S.~J. LaMonte, A.~M. Wolf,
  C.~DeSantis, J.~Lortet-Tieulent, K.~Andrews, D.~Manassaram-Baptiste,
  D.~Saslow, R.~A. Smith, O.~W. Brawley, and R.~Wender.
\newblock Breast cancer screening for women at average risk: 2015 guideline
  update from the american cancer society.
\newblock {\em JAMA}, 314(15):1599--1614, 2015.

\bibitem{ribli_detecting_2018}
D.~Ribli, A.~Horv{\'{a}}th, Z.~Unger, P.~Pollner, and I.~Csabai.
\newblock {Detecting and classifying lesions in mammograms with Deep Learning}.
\newblock {\em Scientific Reports}, 8(1):4165, 3 2018.

\bibitem{Rodriguez-Ruiz2019b}
A.~Rodr{\'{i}}guez-Ruiz, E.~Krupinski, J.~J. Mordang, K.~Schilling, S.~H.
  Heywang-K{\"{o}}brunner, I.~Sechopoulos, and R.~M. Mann.
\newblock {Detection of breast cancer with mammography: Effect of an artificial
  intelligence support system}.
\newblock {\em Radiology}, 2019.

\bibitem{Rodriguez-Ruiz2019a}
A.~Rodriguez-Ruiz, K.~L{\aa}ng, A.~Gubern-Merida, M.~Broeders, G.~Gennaro,
  P.~Clauser, T.~H. Helbich, M.~Chevalier, T.~Tan, T.~Mertelmeier, M.~G.
  Wallis, I.~Andersson, S.~Zackrisson, R.~M. Mann, and I.~Sechopoulos.
\newblock {Stand-Alone Artificial Intelligence for Breast Cancer Detection in
  Mammography: Comparison With 101 Radiologists}.
\newblock {\em JNCI: Journal of the National Cancer Institute}, 2019.

\bibitem{Rosenberg2006}
R.~D. Rosenberg, B.~C. Yankaskas, L.~A. Abraham, E.~A. Sickles, C.~D. Lehman,
  B.~M. Geller, P.~A. Carney, K.~Kerlikowske, D.~S. Buist, D.~L. Weaver, W.~E.
  Barlow, and R.~Ballard-Barbash.
\newblock {Performance benchmarks for screening mammography}.
\newblock {\em Radiology}, 241(1):55--66, 2006.

\bibitem{Saarenmaa1999}
I.~Saarenmaa, T.~Salminen, U.~Geiger, K.~Holli, J.~Isola,
  A.~K{\"{a}}rkk{\"{a}}inen, J.~Pakkanen, A.~Piironen, A.~Salo, and M.~Hakama.
\newblock {The visibility of cancer on earlier mammograms in a population-based
  screening programme}.
\newblock {\em European Journal of Cancer}, 35(7):1118--1122, 7 1999.

\bibitem{Seely2018}
J.~M. Seely and T.~Alhassan.
\newblock {Screening for breast cancer in 2018-what should we be doing today?}
\newblock {\em Current oncology (Toronto, Ont.)}, 25(Suppl 1):S115--S124, 6
  2018.

\bibitem{Sickles2013}
E.~Sickles, C.~D'Orsi, and L.~Bassett.
\newblock {ACR BI-RADS Mammography}.
\newblock In {\em ACR BI-RADS Atlas, Breast Imaging Reporting and Data System,
  5th Edition}, pages 46--74. American College of Radiology, Reston, VA, 2013.

\bibitem{Siu2016}
A.~L. Siu.
\newblock {Screening for breast cancer: U.S. Preventive services task force
  recommendation statement}.
\newblock {\em Annals of Internal Medicine}, 164(4):279--296, 2016.

\bibitem{Szegedy2013}
C.~Szegedy, W.~Zaremba, I.~Sutskever, J.~Bruna, D.~Erhan, I.~Goodfellow, and
  R.~Fergus.
\newblock {Intriguing properties of neural networks}.
\newblock {\em arXiv}, 2013.

\bibitem{Tchou_2010}
P.~M. Tchou, T.~M. Haygood, E.~N. Atkinson, T.~W. Stephens, P.~L. Davis, E.~M.
  Arribas, W.~R. Geiser, and G.~J. Whitman.
\newblock Interpretation time of computer-aided detection at screening
  mammography.
\newblock {\em Radiology}, 257(1):40--46, 2010.

\bibitem{Weedon-Fekjr2008}
H.~Weedon-Fekj{\ae}r, B.~H. Lindqvist, L.~J. Vatten, O.~O. Aalen, and
  S.~Tretli.
\newblock {Breast cancer tumor growth estimated through mammography screening
  data}.
\newblock {\em Breast Cancer Research}, 10(3):R41, 2008.

\bibitem{WHO2014}
{World Health Organization}.
\newblock {WHO position paper on mammography screening}.
\newblock \url{www.who.int/cancer/publications/mammography_screening/en}, 2014.

\bibitem{Wu2019ValidationRates}
K.~Wu, E.~Wu, Y.~Wu, H.~Tan, G.~Sorensen, M.~Wang, and B.~Lotter.
\newblock {Validation of a deep learning mammography model in a population with
  low screening rates}.
\newblock In {\em Fair ML for Health Workshop}. Neural information processing
  systems foundation, 2019.

\bibitem{Wu2019}
N.~Wu, J.~Phang, J.~Park, Y.~Shen, Z.~Huang, M.~Zorin, S.~Jastrzebski,
  T.~Fevry, J.~Katsnelson, E.~Kim, S.~Wolfson, U.~Parikh, S.~Gaddam, L.~L.~Y.
  Lin, K.~Ho, J.~D. Weinstein, B.~Reig, Y.~Gao, H.~T.~K. Pysarenko, A.~Lewin,
  J.~Lee, K.~Airola, E.~Mema, S.~Chung, E.~Hwang, N.~Samreen, S.~G. Kim,
  L.~Heacock, L.~Moy, K.~Cho, and K.~J. Geras.
\newblock {Deep Neural Networks Improve Radiologists' Performance in Breast
  Cancer Screening.}
\newblock {\em IEEE transactions on medical imaging}, 10 2019.

\bibitem{Yala2019a}
A.~Yala, C.~Lehman, T.~Schuster, T.~Portnoi, and R.~Barzilay.
\newblock {A deep learning mammography-based model for improved breast cancer
  risk prediction}.
\newblock {\em Radiology}, 2019.

\bibitem{Yala2019}
A.~Yala, T.~Schuster, R.~Miles, R.~Barzilay, and C.~Lehman.
\newblock {A Deep Learning Model to Triage Screening Mammograms: A Simulation
  Study}.
\newblock {\em Radiology}, 2019.

\bibitem{zech2018variable}
J.~R. Zech, M.~A. Badgeley, M.~Liu, A.~B. Costa, J.~J. Titano, and E.~K.
  Oermann.
\newblock Variable generalization performance of a deep learning model to
  detect pneumonia in chest radiographs: A cross-sectional study.
\newblock {\em PLoS medicine}, 15(11):e1002683, 2018.

\end{thebibliography}

\bibliographystyle{ieee}

\end{document}